\documentclass[11pt,twoside]{article}

\usepackage{asp2006}
\usepackage{epsf}
\usepackage{psfig}
\usepackage{lscape}
\usepackage{natbib}
\usepackage{graphicx}
\usepackage{color}
\usepackage{setspace}

\begin{document}
\title{The small-scale solar surface dynamo}   
\author{Jonathan {Pietarila Graham}, Sanja Danilovic, and Manfred Sch\"ussler}   
\affil{Max-Planck-Institut f\"ur Sonnensystemforschung, 37191 Katlenburg-Lindau, Germany}    

\begin{abstract} 
The existence of a turbulent small-scale solar surface dynamo is likely,
considering existing numerical and laboratory experiments, as well as comparisons
of  a small-scale dynamo in MURaM
simulations with {\sl Hinode}
observations.  We find the
observed peaked probability distribution function (PDF) from
Stokes-$V$ magnetograms is consistent with a monotonic PDF of the actual vertical
field strength.  The cancellation function of the
vertical flux density from a {\sl Hinode} SP observation is found to
follow a self-similar power law over two decades in length scales down
to the $\approx200\,$km resolution limit.  This provides observational
evidence that the scales of magnetic structuring in the photosphere
extend at least down to $20\,$km.  From the power law, we
determine a lower bound for the true quiet-Sun mean vertical unsigned
flux density of $\approx43\,$G, consistent with our
numerically-based estimates that 80\% or more of the vertical unsigned
flux should be invisible to Stokes$-V$ observations at a resolution of
$200\,$km owing to cancellation.  Our estimates significantly reduce the order-of-magnitude
discrepancy between Zeeman- and Hanle-based estimates.
\end{abstract}



\section{Introduction}

The existence of a global solar dynamo is generally accepted, though
it is not well understood.  There is evidence for a
second, local dynamo operating near the solar surface -- the
so-called solar-surface dynamo \citep{PeSz1993}.\footnote{Actually, in
simulations, turbulent dynamo action occurs in the bulk of the convection
zone, though greater than solar rotation rate may be required to generate
large-scale field in the absence of shear \citep{BrBrBr+2007}.}
  This
dynamo may be driven by turbulence:  magnetic energy
is amplified by random stretching of magnetic field lines by
turbulent motions \citep{Ba1950}. 
Magnetic energy is
lost to Ohmic dissipation. For dynamo action to
succeed, stretching must dominate over dissipation.
The relative strengths of these two effects is quantified by the
magnetic Reynolds number, $Re_M \equiv v_0l_0/\eta$ ($v_0$ and
$l_0$ are typical velocity and length scales and $\eta$
the magnetic diffusivity).  Similarly, the kinetic Reynolds number is
$Re \equiv v_0l_0/\nu$ where $\nu$ is the kinematic
viscosity.  When $Re_M$ exceeds a critical threshold, $Re_M >
Re_M^C$, dynamo action results.

Two different types of dynamo can be found, depending on the presence
or absence of net flow helicity \citep{MeFrPo1981}.  With net helicity, magnetic energy grows at scales
larger than the energy-containing scale of the fluid motions: large-scale
dynamo (LSD) or mean-field dynamo.  LSDs are often studied with
mean-field theory; the production of large-scale magnetic
energy is approximately the alpha-effect (see, e.g., \citealt{Br2003}).  Without
net helicity, dynamo action is harder to
achieve and magnetic every grows at scales smaller than the forcing
scale.  This latter defines small-scale dynamo (SSD) or fluctuation
dynamo action.  Near the solar surface, the convective time scale is
much shorter than the rotation period, the effects of rotation can
be neglected, and a flow with no net helicity results.  Any surface dynamo
will thus be a SSD.

This is also suggested by observation of the small-scale
magnetic field in the quiet Sun.  In high resolution magnetograms we
see mixed polarity fields on small scales, which is variously
called the magnetic carpet or the salt-and-pepper pattern
\citep{TiSc1998,HaScTi2003}, consistent with the idea
of SSD action.  As turbulent
convection can drive small-scale dynamo action (in numerical
simulations of Boussinesq convection without rotation
\citealt{C99,CaEmWe2003}), observations and simulations together
provide evidence of a SSD
driven by turbulence at the solar surface (likely deeper as well).

Several arguments can be put forward against a small-scale solar dynamo.  Firstly, there exists the possibility that the
small-scale field is solely produced by the
shredding up of large-scale field by turbulence.  However,
observationally the amount of
small-scale flux is not dependent on the solar cycle
\citep{HaScTi2003,TrBuShAsRa2004}. This might not be the case if the
small-scale flux is the result of the shredding up of the field from
the global dynamo.  It might also then show some latitudinal
dependence (among low latitudes).  Assuming the existence
of both SSD and shredding,
 small-scale
dynamo is predicted to create small-scale magnetic field at the
turbulent rate of stretching ($\propto Re_M^{1/2}$, see, e.g., \citealt{IsScCo+2007})
 which is much faster either than
large-scale field can be produced (at time scales associated with the
kinetic-energy-containing length scales) or be shredded up by the
turbulence \citep{ScHaBr+2005}.

Secondly, small-scale dynamo
action may not be possible at the magnetic Prandtl number, $P_M$, of
the solar plasma, $P_M \equiv Re_M/Re \approx 10^{-5}$:
$Re_M^C$ sharply increases with decreasing $P_M$ (increasing $Re$)
since eddies smaller than the characteristic scale of the magnetic field diffuse the
field and inhibit dynamo action.  The two asymptotic possibilities are
$Re_M^C\rightarrow\,$const as $Re\rightarrow\infty$ (SSD at low $P_M$;
\citealt{RoKl1997,BoCa2004}) or $Re_M^C/Re=P_M^C\rightarrow\,$const as
$Re\rightarrow\infty$ (no SSD at low $P_M$; \citealt{ScHaBr+2005}).
Numerical simulations of low $P_M$ SSDs have focused on the
existence (or not) of a time-averaged mean flow.  Such a flow
exists for the Sun and many other astrophysical cases.  Both
with \citep{PMM+05} and without a mean flow \citep{IsScCo+2007}, a
plateau in $Re_M^C$ was found.  This suggests such dynamo
action should be possible on the Sun.  However, the $P_M$ of the sun
is 3 orders of magnitude smaller that that accessible to present numerical
computations; the observed plateaus may not represent the asymptotic
behavior.  Fortunately, a laboratory dynamo resulting from unconstrained
turbulence in liquid sodium ($P_M\approx10^{-5}$) has been
demonstrated \citep{MoBeBo+2007} establishing that a turbulent
dynamo is possible at values of $P_M$ corresponding to the solar plasma.

\section{The MURaM code and the solar surface dynamo}

The remaining objection against a surface dynamo was raised by
\citet{StBeNo2003}: unlike the \citet{C99} simulation, the Sun is
strongly stratified and magnetic flux is swept into the down-flow
lanes and subject to long recirculation times.  In the
simulations of \citet{StBeNo2003} with open
boundaries, no dynamo action was found.  However, \citet{VoSc2007}
have demonstrated SSD action for sufficiently high
$Re_M$ using the MURaM code \citep{V03,VSS+05}.  
They simulated a box with strong stratification
and open boundaries, but prevented advection of magnetic flux from outside
the box: providing
an artificially isolated surface layer ignoring any SSD
action that may occur in deeper layers.  They conducted 3
simulations with increasing
$Re_M$ (see Table \ref{table:compare}).  Above the critical threshold,
$Re_M^C\sim2000$, dynamo action occurs.  The magnetic energy
spectrum peaks at scales smaller than the energy-containing scale
of the fluid motions, demonstrating that SSD
{\sl is} a possibility for a solar surface dynamo
despite strong stratification and little recirculation. In Table~\ref{table:compare},
dynamo action in the 3 MURaM simulations of \citet{VoSc2007}
as well as in the simulations of \citet{C99} and of \citet{StBeNo2003} is
indicated solely by their $Re_M$ and $P_M$.  For the simulations
below the critical threshold for dynamo action, ($Re_M^C\ga1000$),
there is no dynamo action.  The result by \citet{C99} is an exception
as $P_M>1$ and this lowers the threshold.

\begin{table}[!ht]
\caption{Simulations by \citet{C99},
  \citet{StBeNo2003}, and \citet{VoSc2007}:
  computational grid size, magnetic Reynolds and Prandtl
  numbers, boundary conditions (BC; open or closed),
  and presence of SSD.}
\smallskip
\begin{center}
{\small
\begin{tabular}{lccccc}
\tableline
\noalign{\smallskip}
{Simulation} & {Grid pts.} & {$Re_M$} & {$P_M$} & {BC} & {SSD}\\
\noalign{\smallskip}
\tableline
\noalign{\smallskip}
      {Run A} & {$288^2\times100$} & {$\approx 300$} & {$\la 1$} & {open} & {N}\\
      {Stein et al.} & {$253^2\times163$} & {$\approx 600$} & {$\approx 1$} & {open} & {N}\\ 
      {Run B} & {$576^2\times100$} & {$\approx 1300$} & {$\la 1$} & {open} & {N}\\
      {Cattaneo} & {$512^2\times97$} & {$\approx 1000$} & {$\approx 5$} & {closed} & {Y}\\ 
      {Run C} & {$648^2\times140$} & {$\approx 2600$} & {$\la 1$} & {open} & {Y}\\
\noalign{\smallskip}
\tableline
    \end{tabular}
}
  \end{center}
\label{table:compare}
\end{table}

The MURaM Run C dynamo also reproduces, quantitatively, some aspects
of observations.  For instance, \citet{LiKuSoNa+2008} find
from {\sl Hinode} SP observations that
the mean horizontal magnetic field is stronger than the mean unsigned
vertical magnetic field.  Using a spatial map, 
in the ``normal mode'' 
($324^{''}\times 164^{''}$), they make magnetograms of the apparent
longitudinal flux density, $B_{\mbox{app}}^L$, and apparent transverse
flux density, $B_{\mbox{app}}^T$.  They find a factor of 5 stronger
horizontal fields: the means are
$\left\langle|B_{\mbox{app}}^L|\right\rangle\approx 11\,$G and
$\left\langle B_{\mbox{app}}^T\right\rangle\approx 55\,$G.  A similar ratio was
found in the MURaM Run C dynamo by \citet{ScVo2008}; see also
\citet{StReSc+2008}.  In the MURaM simulation, both
small-scale, low-lying vertical loops and extended
canopy-like structures are found in the line formation region.  These
structures (related to flux expulsion of horizontal field)
contribute to the stronger mean horizontal fields:
horizontal field occupies a larger area than the
narrow vertical foot points.  The simulation also makes a prediction:
lines with different formation heights will yield different
ratios.

\section{Measurement of the turbulent magnetic field}

\begin{figure}[!ht]
\begin{center}
  \includegraphics[width=8.125cm]{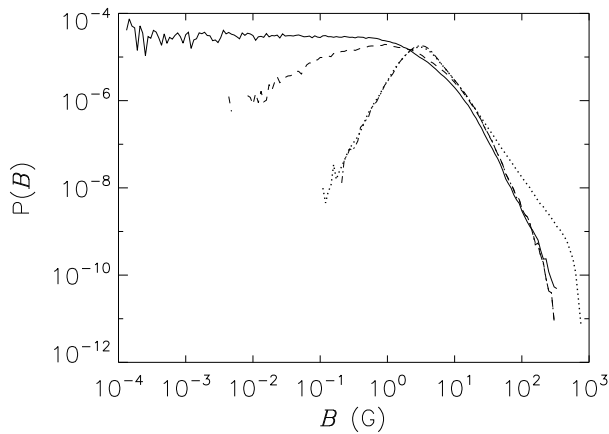}
\end{center}
  \caption{PDFs for magnetic
    field strengths and derived proxies: MURaM
    simulation $B_{\mbox{ave}}$ (solid line), {\sl Hinode} SP
    observation {$B^{\mbox{L}}_{\mbox{app}}$} (dotted line), and synthetic
    {$B^{\mbox{L}}_{\mbox{app}}$} from MURaM (dashed) -- including noise
    (dot-dashed).}
  \label{FIG:PDFS}
\end{figure}

Whatever its source, the small-scale quiet-Sun magnetic field is
turbulent.  This should be taken into consideration when interpreting
observations
\citep{SaAlLadeMaPi+1996,SaAlLi2000}.
Consider the discrepancy between the probability
distribution functions (PDFs) of the apparent vertical flux densities
from the {\sl Hinode} SP observation (discussed in the previous section)
and that of $B_z$ in MURaM dynamo Run C
(see Fig. \ref{FIG:PDFS}).  For the SP observation, the PDF is peaked
at $\approx3\,$G.  The simulation PDF is
monotonic; there is a much greater amount of weak field than in
observations.  Here $B_{\mbox{ave}}$ is defined as $B_z$ averaged over the height range corresponding
to $\log\tau\in[-3.5,.1]$: other samplings show similar PDFs.  Synthetic \texttt{FeI} $630\,$nm doublet spectra are generated
from the MURaM cube and used to create magnetogram signals.
The PDF of the result is shown as a dashed line in Fig. \ref{FIG:PDFS}
and begins to resemble the peaked PDF of the SP observation.
With the addition of instrumental noise the PDF  (dash-dotted line) becomes quite similar to
that of the {\sl Hinode} observation.  The observation has
a much greater amount of strong field due to the much
higher $Re_M$ of the Sun (more efficient SSD)
and the presence of network elements which are not in the simulation.
In any case, these considerations show that
the observed peaked PDF is in fact consistent with a much greater
prevalence of weak vertical magnetic field.

\begin{figure}[!ht]
\begin{center}
  \begin{tabular}{c@{\hspace{.15in}}c}
  \includegraphics[width=6.6cm]{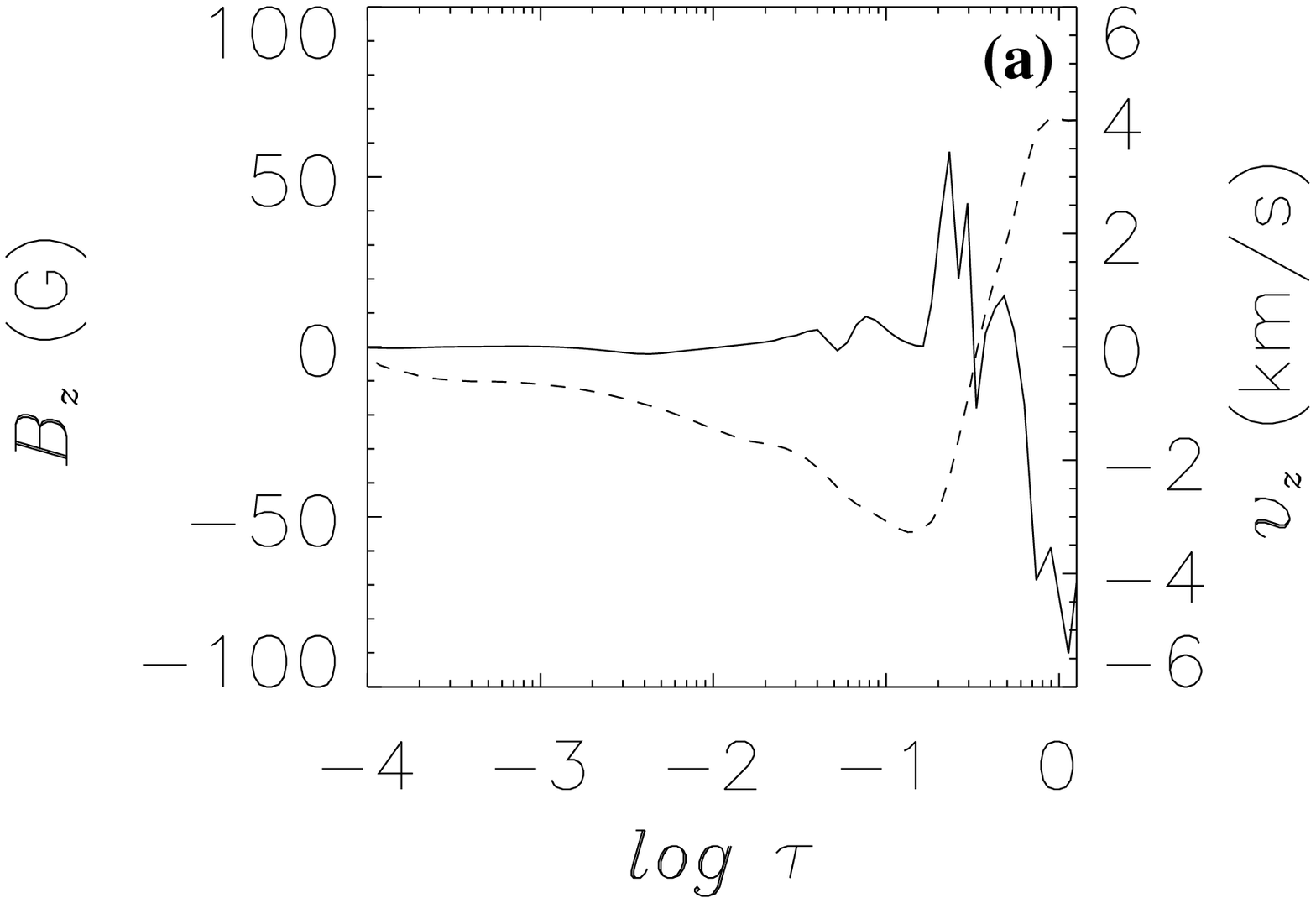} &
  \includegraphics[width=6.6cm]{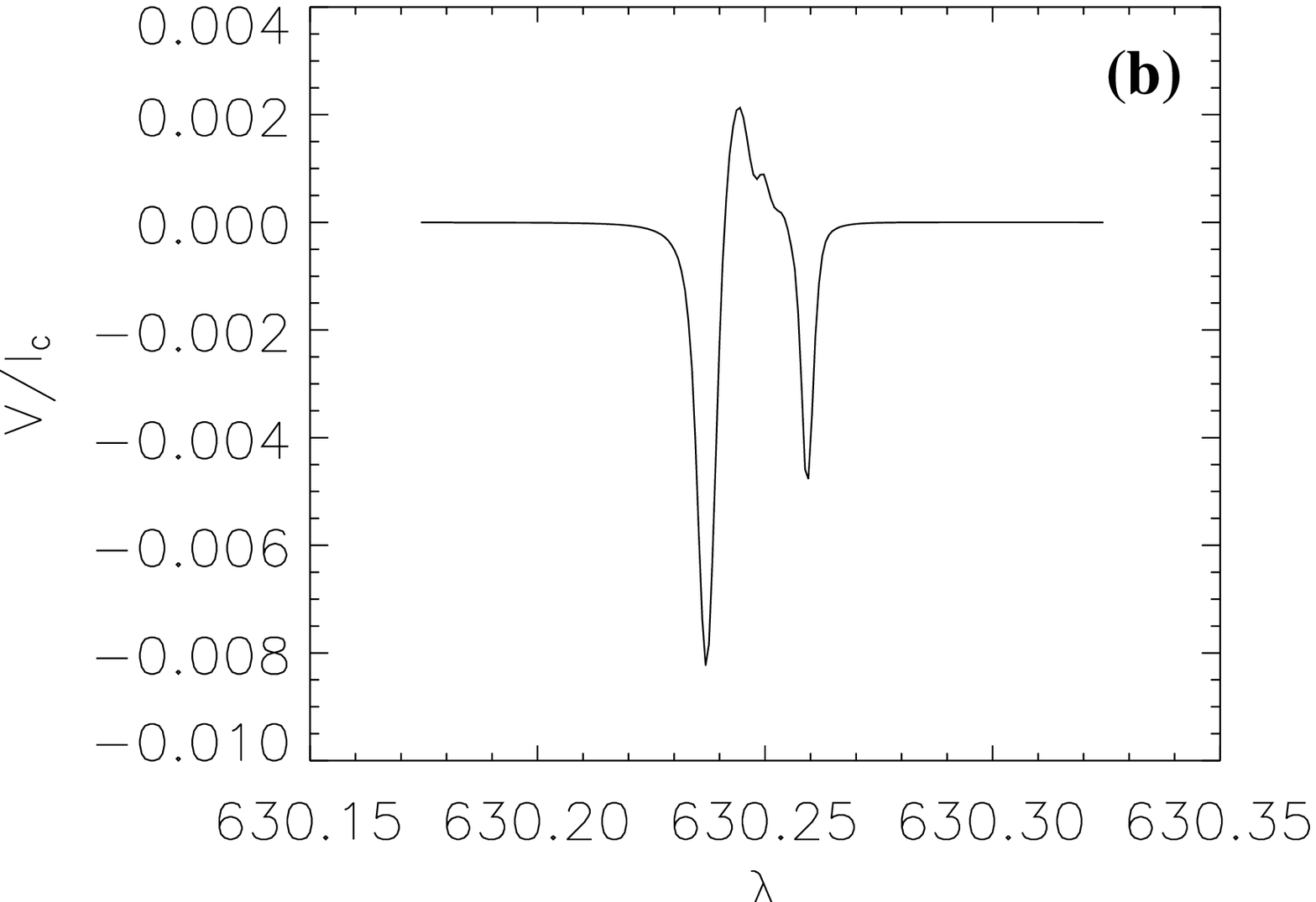}  \end{tabular}
\end{center}
  \caption{{\bf (a)} $B_z$ (solid line) and $v_z$ (dashed line)
    versus  $\tau_{500\,\mbox{nm}}$, and {\bf (b)}
    Stokes~$V$ profile for a pixel with
    $B^{\mbox{L}}_{\mbox{app}}=-9\,$G and
    $B_{\mbox{ave}}=-5\cdot10^{-3}\,$G.  Strong gradients can lead to
    $|B^{\mbox{L}}_{\mbox{app}}| \gg |B_{\mbox{ave}}|$.}
  \label{FIG:RAY}
\end{figure}

\begin{figure}[!ht]
\begin{center}
\includegraphics[width=8.125cm]{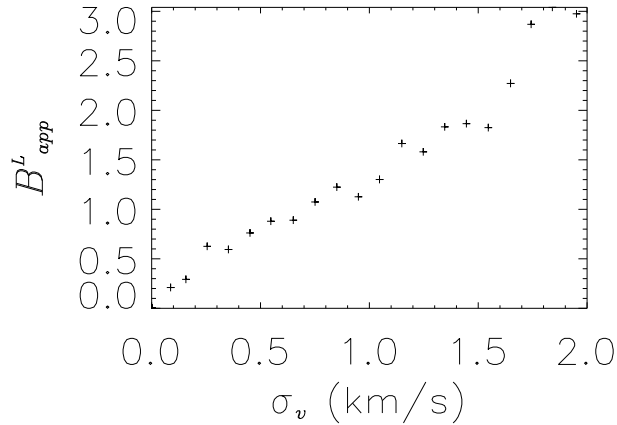}
\end{center}
  \caption{Average $B^{\mbox{L}}_{\mbox{app}}$ versus
    the fluctuations of the vertical velocity along the
    line-of-sight, $\sigma_v$ ($|B_{\mbox{ave}}|<0.1\,$G for all pixels).}
  \label{FIG:SAMI}
\end{figure}

Vertical radiative transfer through a
turbulent fluid contributes to this deformation of the PDF
 \citep{SaAlEmCa2003}.  In Fig.~\ref{FIG:RAY}(a), we plot
the $z-$components of the magnetic field and velocity for a simulation pixel with $B_{\mbox{ave}}\ll B^{\mbox{L}}_{\mbox{app}}$.  The
magnetic field undergoes a reversal above $\tau=1$.  The average of
$B_z$ over the line formation region is very small ($\sim10^{-3}\,$G),
but the reversal of $B_z$ is coincident with a strong gradient in
$v_z$.  Thus, the absorption profiles from the two
opposite-polarity fields are Doppler shifted with respect to each
other and do not cancel. Such cancellation leads to a stronger Stokes $V$ signal
than would otherwise arise (also an asymmetric one, see
Fig. \ref{FIG:RAY}(b)).  If this statistically accounts for the
differences in the PDFs, a correspondence
between the strength of the velocity fluctuations along the line of
sight and the synthetic magnetogram would be expected (for pixels with weak
$B_{\mbox{ave}}$).  This is indeed seen in Fig. \ref{FIG:SAMI}, which
includes only pixels with $|B_{\mbox{ave}}|<0.1\,$G. There
exists a definite trend of stronger magnetogram signal when
there are stronger Doppler shifts (velocity gradients) between the various layers of the
atmosphere.

\begin{figure}[!ht]
\begin{center}
  \begin{tabular}{c@{\hspace{.15in}}c}
  \includegraphics[width=6.6cm]{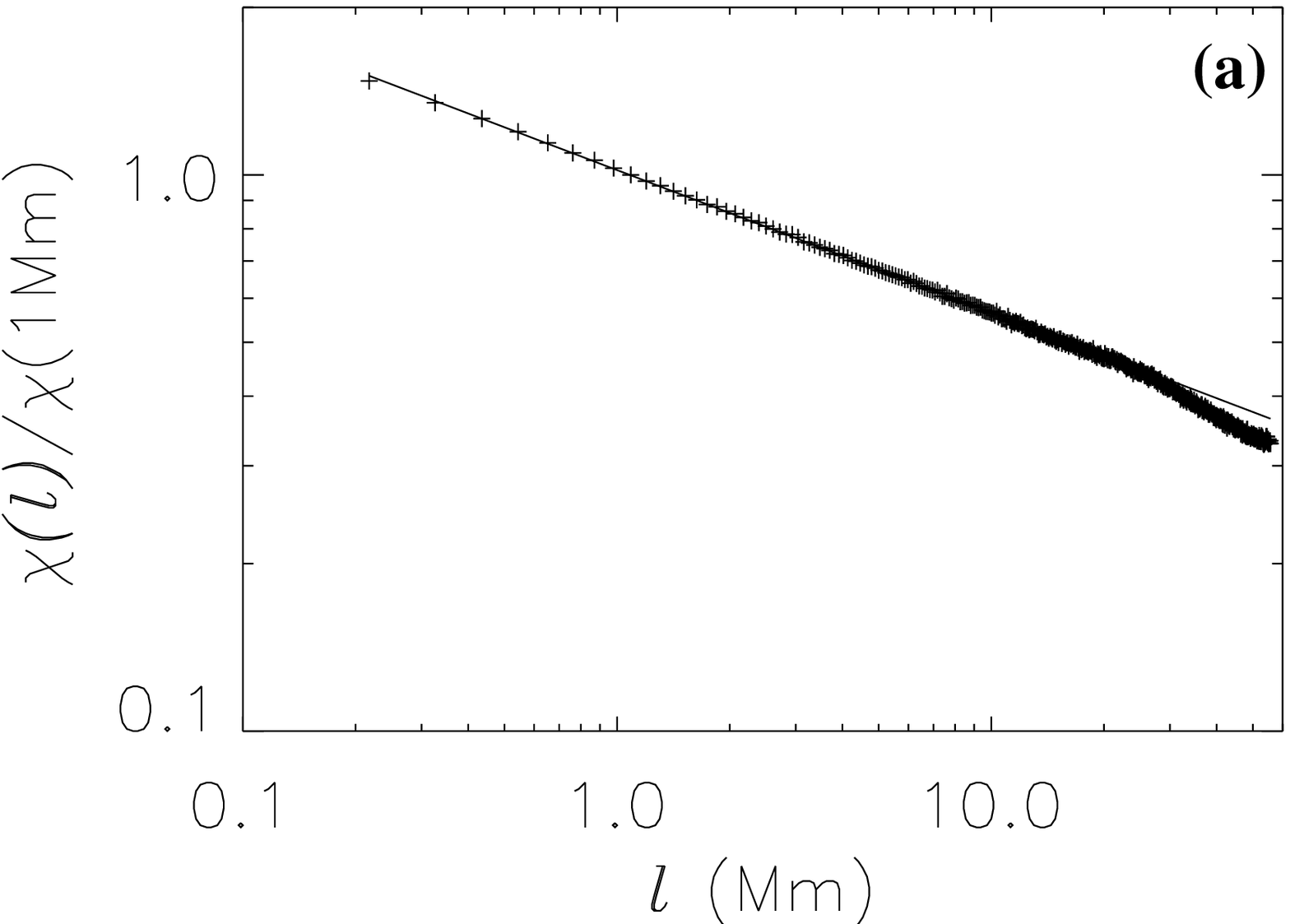} &
  \includegraphics[width=6.6cm]{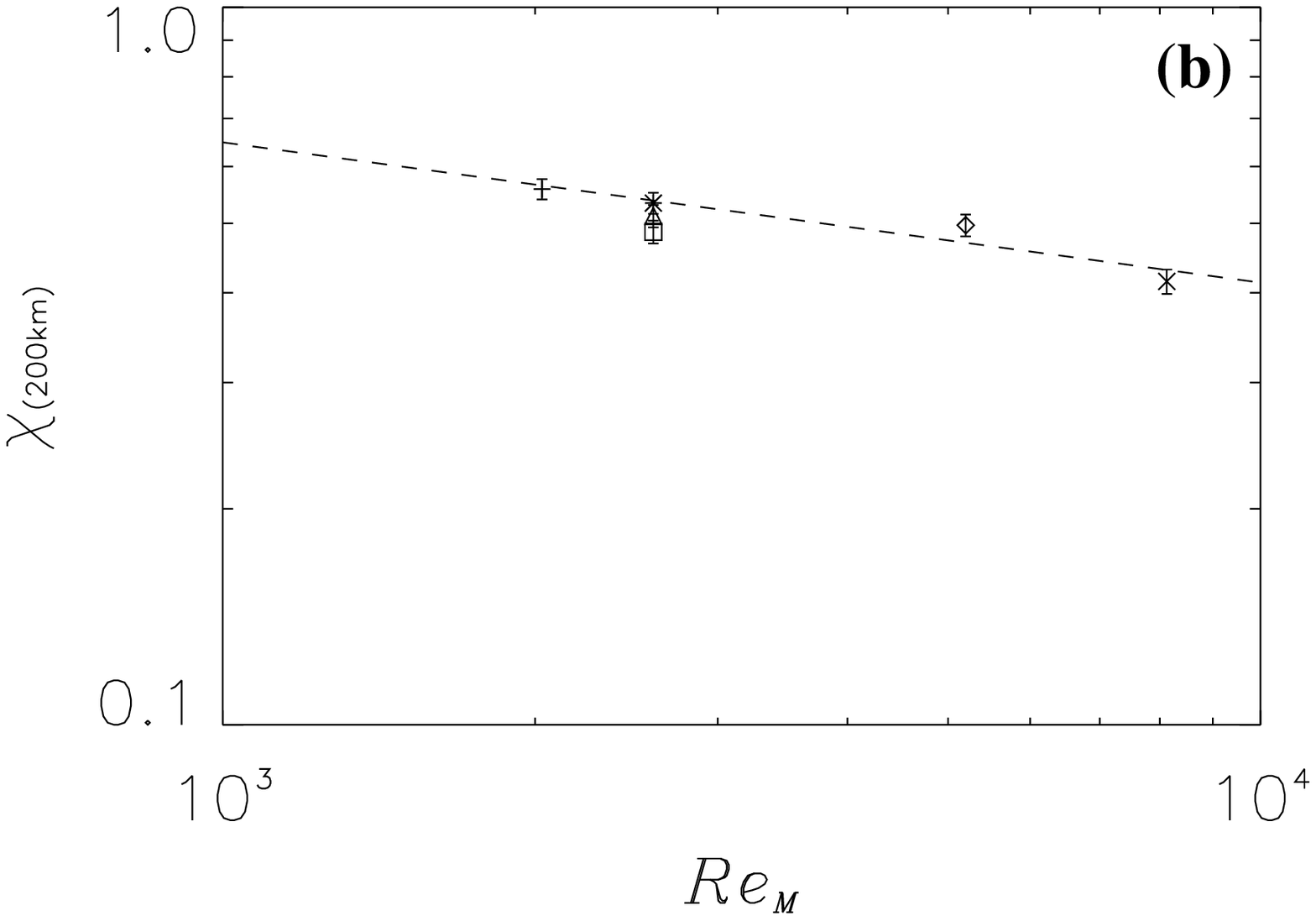}  \end{tabular}
\end{center}
  \caption{{\bf(a)} Normalized cancellation function,
    $\chi(l)/\chi(1\,$Mm$)$, versus scale, $l$, from {\sl Hinode}
    $B^{\mbox{L}}_{\mbox{app}}$.  The exponent of the fitted power law
    is $\kappa=0.26 \pm 0.01$.
    {\bf(b)} Portion of flux remaining at $l=200\,$km,
    $\chi(200\,$km$)$, versus $Re_M$ for MURaM dynamo.
A
   power-law is indicated by the
    dashed line.}
  \label{FIG:HINODE_CAEXP}
\end{figure}

How do gradients in the horizontal direction affect observations?  Or,
given that opposite-polarity vertical field inside a
resolution element leads to cancellation of Stokes $V$ signal, how can
the turbulent (and hence fractal) nature of the magnetic field be used
to estimate the cancellation (i.e., to get an estimate for the true
unsigned vertical flux density)?  Fractals are self-similar; they
display power-law scaling such as in the box-counting fractal
dimension: the domain is partitioned into boxes of edge length
$l$ and the number of boxes containing the fractal, $N(l)$, has a
power-law relation, $N(l)\propto~l^{-D_f}$, where
$D_f$ is the fractal dimension.  The magnetic field,
however, is more complex; for each box in a
partition, there is a net flux (a magnitude) and a direction (sign),
either up or down.  In this case, we need to use the cancellation
function,
\begin{equation}
\chi(l) \equiv {\sum_i \bigg{|}\int_{\mathcal{A}_i(l)}B_zda \bigg{|}}
      \bigg{/}{\int_{\mathcal{A}}|B_z|da}
\end{equation}
introduced by \citet{ODS+92}.  Simply put, $\chi(l)$,
measures the portion of net flux remaining after averaging over boxes of
length $l$.
A fractal (self-similar) magnetic field has
a power-law scaling, $\chi(l)\sim l^{-\kappa}$. For the
magnetogram from the {\sl Hinode} SP observation (see
Fig. \ref{FIG:HINODE_CAEXP} (a)), a clear power-law scaling of
the cancellation function exists over two decades of length scales down
to the resolution limit at $\approx200\,$km.  Due to experience with
turbulent power-laws, we can be confident that more
cancellation occurs at scales down to $\approx20\,$km or less
as the power-law continues down to the resolution limit
\citep{CaBr1997,SVCV+04}. It is unknown how far this
power law extends, but it must stop before the magnetic
dissipation scale, $l_\eta$.  Assuming that the power law
continues until dissipation sets in, we derive an estimate for the
true, mean unsigned vertical magnetic field for the Sun, $\langle|B_z|\rangle$,
defined as
\begin{eqnarray}
\langle|B_z|\rangle \equiv  {\int_{\mathcal{A}}\bigg{|}B_z\bigg{|}da }
  \bigg{/}{\int_{\mathcal{A}}da}\,.
\end{eqnarray}
The
mean absolute value of $B_z$ measured at resolution $l$ is
\begin{eqnarray}
\langle|B_z|\rangle_l \equiv {\sum_i \bigg{|}\int_{\mathcal{A}_i(l)}B_zda \bigg{|}}
\bigg{/} \int_{\mathcal{A}}da\,.
\end{eqnarray}
Note that $\langle|B_z|\rangle_l = \chi(l) \cdot \langle|B_z|\rangle$.
Using that there is no cancellation below the magnetic dissipation scale,
$\langle|B_z|\rangle_{l_\eta}=\langle|B_z|\rangle$,
\begin{eqnarray}
\langle|B_z|\rangle =  \langle|B_z|\rangle_l \cdot
\frac{\chi(l_{\eta})}{\chi(l)}
= 12\mbox{G}   \cdot
\left(\frac{\bf 100\,\mbox{km}}{\bf l_{\eta}}\right)^{\bf 0.26}
\label{eq:extrap}
\end{eqnarray}
where $12\,$G is the $100\,$km result from \citet{LiKuSoNa+2008}. 
We apply Kolmogorov theory to estimate $l_\eta$ for the Sun.
Assuming $\eta \sim 10^8\,\mbox{cm}^2\mbox{s}^{-1}$ \citep{KoCr1983},  we find $l_\eta$ to be $\sim80\,$m.
At this scale, $\chi(l)$ must have a slope of zero and the power-law is
affected for about a decade of scales larger.  There
will still be some cancellation but the slope will be
decreasing.  Therefore, our calculation is based on a scale of
$800\,$m in Eq. (\ref{eq:extrap}), giving a lower bound of
$\approx43\,$G.

An estimation for $\langle|B_z|\rangle$ can also be obtained using the MURaM
dynamo simulation.  In Fig. \ref{FIG:HINODE_CAEXP} (b), we plot the
cancellation at a resolution of $200\,$km
versus $Re_M$.  The power-law scaling can be
extrapolated to the expected $Re_M$ of the Sun $\sim3\cdot10^5$:
we
    estimate $\chi(200km)\la1/5$,
which means $\left\langle|B_z|\right\rangle\ga11\cdot5\sim55\,$G.  This provides qualitative
agreement with Hanle estimations.  Based on our estimates, the mean
unsigned vertical field strength is about $40$ or $50\,$G.  As shown
by \citet{LiKuSoNa+2008}, the mean horizontal
field strength is stronger. Therefore, the mean vector magnitude of
the magnetic field of $\sim130\,$G as reported by
\citet{TrBuShAsRa2004} is not in contradiction with these Zeeman results.

\section{Summary}

A small-scale solar dynamo near the surface is likely.  MURaM simulations show
dynamo action and its properties are found to be in
agreement with {\sl Hinode} observations.  The arguments against a solar small-scale dynamo,
thus far, have failed.  Whatever its source, the
small-scale magnetic field is turbulent and fractal and this should be
taken into consideration when interpreting observations: 1)  the PDF of Stokes $V$ field estimates do not accurately
represent the PDF of the $B_z$ and are
consistent with a much greater prevalence of weak $B_z$; 2) the multi-fractal
self-similar pattern of the quiet-Sun photospheric magnetic field
extends down to the resolution limit,
$200\,$km.  This constitutes observational evidence that the
smallest scale of magnetic structuring in the photosphere is $20\,$km
or smaller.  The power law also constrains the quiet-Sun
true mean unsigned vertical flux density: the lower bound,
$\approx43\,$G, is consistent with
estimates based solely on numerical
simulations  ($\sim55\,$G).
The order of magnitude disparity between Hanle and
Zeeman-based estimates may be resolved by a proper consideration
of the cancellation properties of the full vector field.

\acknowledgements The authors would like to acknowledge fruitful
discussions with S. Solanki, A. Pietarila, and R. Cameron.  {\sl
  Hinode} is a Japanese mission developed and launched by ISAS/JAXA,
with NAOJ as domestic partner and NASA and STFC (UK) as international
partners. It is operated by these agencies in co-operation with ESA
and NSC (Norway).



\end{document}